\documentclass[modern]{aastex61}
\usepackage{graphicx}
\usepackage{amsmath}
\usepackage{amssymb}
\usepackage{enumitem}

\newcommand\mybar{\kern1pt\rule[-\dp\strutbox]{.8pt}{\baselineskip}\kern1pt}

\setlist[itemize]{noitemsep, topsep=0pt, leftmargin=*}

\shorttitle{Localization of CNEOS 2014-01-08}
\shortauthors{Loeb}

\begin{document}

\title{Peak-Brightness Localization of the CNEOS 2014-01-08 (IM1) Fireball}

\author{Abraham Loeb}
\affiliation{Astronomy Department, Harvard University, 60 Garden
  St., Cambridge, MA 02138, USA}

\begin{abstract}
In a recent preprint, Fernando et al. (2024) used public data from
infrasound stations to constrain the localization of the fireball of
the CNEOS 2014-01-08 (IM1) bolide. The analysis inferred a
90\%-confidence ellipse with semi-minor and semi-major axes of 186 and
388 km, respectively. This large error ellipse includes the much better
localization box derived by sensors aboard U.S. Government satellites
which detected the fireball light. At the fireball's peak brightness,
the CNEOS localization box documented by NASA/JPL measures 11.112~km on
a side and is centered on a latitude of 1.3$^\circ$S and a longitude
of 147.6$^\circ$E. Here, we point out that the recent expedition to
retrieve materials from IM1's site (Loeb et al. 2024a,b,c) surveyed a
region of tens of km around the CNEOS box center, and was not dictated
by the data studied by Fernando et al. (2024) because of its larger
uncertainties.

\end{abstract}

\section{Introduction}
On January 8, 2014, US government satellites detected the light from a
fireball of a meteor, labeled CNEOS 2014-01-08 (hereafter abbreviated
as IM1), that was moving significantly faster than the escape speed
from the solar system~\citep{SL22}. The location, velocity and
radiated energy of the meteor were documented in the official
website\footnote{\url{https://cneos.jpl.nasa.gov/fireballs/}} of the
Center for Near Earth Object Studies (CNEOS) fireball catalog,
compiled by NASA/JPL. The interstellar origin of IM1's velocity vector
was double-checked and confirmed at the 99.999\% confidence level in
an official letter dated March 1, 2022 from the US Space Command to
NASA.\footnote{\url{https://lweb.cfa.harvard.edu/~loeb/DoD.pdf}}

The lightcurve of the fireball was included in the CNEOS
database,\footnote{\url{https://cneos.jpl.nasa.gov/fireballs/lc/bolide.2014.008.170534.pdf}}
with a document narrative explaining that the detection was made by
``sensors aboard U.S. Government satellites'', which were triggered by
a ``flash signature of a large meteoroid entry into the atmosphere''
at a latitude of 1.2$^\circ$S and a longitude of 147.1$^\circ$E. The
peak brightness of the fireball was reported in the CNEOS catalog
table at a latitude of 1.3$^\circ$S, a longitude of 147.6$^\circ$E and
an altitude of 18.7~km. The vector connecting the atmospheric entry
flash and the fireball's peak brightness agrees with the measured
direction of motion of the bolide after its atmospheric entry, as
documented in the CNEOS catalog table.

The fireball lightcurve shows three prominent, equally-separated flares, with
the peak brightness associated with the last flare, ending about 0.3~s
after the beginning of the first prominent flare. The reported localization and
altitude in the CNEOS catalog table correspond to the brightest third flare.

\section{Peak-Brightness Localization}
The peak-brightness location of IM1 was defined in the CNEOS fireball
catalog to within a tenth of a degree precision in latitude and
longitude, corresponding to 11.112~km on the Pacific Ocean surface,
centered about 90 km away from Manus Island in Papua New
Guinea. Better localization was not meaningful since the fireball's
speed of 44.8~${\rm km~s^{-1}}$ and direction of motion at a
$31^\circ$ angle relative to the ocean surface, implied that the
bolide traveled across a path of 13.44~km over the 0.3~s duration of
its prominent flares.

On June 14-28, 2023, an expedition was conducted to retrieve
meteoritic materials from IM1's
site~\citep{Loeb24a,Loeb24b,Loeb24c}. The surveyed region centered on
the CNEOS localization box and included 26 runs across a region
extending out to several tens of km away from the box center, as shown
in Figure 2 in \cite{Loeb24a}.

In a new preprint, \citet{Fer24} analyzed seismic and acoustic data
from a wide region around IM'1 impact site. This data is independent
from the data reported by the U.S. Government satellite sensors. Based
on infrasound data, \cite{Fer24} inferred a 90\%-confidence
localization ellipse with semi-minor and semi-major axes of 186 and
388 km, respectively, and an area of 227,000~${\rm km}^2$, which
includes the CNEOS localization box but is centered at a distance of
$\sim 170~{\rm km}$ from it. Whereas this result is consistent with
the much better CNEOS localization, \citet{Fer24} suggest that the
expedition might have been misguided in surveying the region
identified by the U.S. Government satellites based on the detected
fireball light.

\section{Discussion}
\cite{LS2} considered seismometer data from Manus Island (AU.MANU) to
test consistency with the CNEOS localization box, altitude and
timing. However, the time delay associated with any acoustic or
seismic signal detected by a single station, can only constrain
within uncertainties the distance of the fireball from that station,
implying a fireball location within a circular band of some
uncertainty-width in all possible directions around the
station. Consequently, the region surveyed by the expedition (Figure 2
in \citet{Loeb24a}) could not have been defined by this analysis, as
suggested by \cite{Fer24}. Actual localization requires multiple
stations as used by \cite{Fer24}, but that analysis resulted in a
large 90\%-confidence ellipse - consistent with the better localization
box derived by U.S. Government satellites and documented by NASA/JPL
in the CNEOS catalog.

\bigskip
\bigskip
\section*{Acknowledgements}

This work was supported in part by Harvard's {\it Galileo Project}.

\bigskip
\bigskip
\bigskip

\bibliographystyle{aasjournal}
\bibliography{m}
\label{lastpage}
\end{document}